\documentclass{LT23auth}
\usepackage{graphicx}

\begin{document}

\begin{frontmatter}

  \title{Comparing measured and calculated local density of states in
    a disordered two-dimensional electron system}

\author[address1]{M. Morgenstern}, 
\author[address1]{J. Klijn}, 
\author[address1]{Chr. Meyer}, 
\author[address2]{R. A. R\"omer\thanksref{thank1}},
\author[address1]{R. Wiesendanger}

\address[address1]{Institute of Applied Physics, Hamburg University,
  Jungiusstra\ss e 11, D--20355 Hamburg, Germany}

\address[address2]{Institute of Physics, Chemnitz University of
  Technology, 09107 Chemnitz, Germany}

\thanks[thank1]{Corresponding author. Permanent address: Department of
  Physics, University of Warwick, Coventry CV4 7AL, UK, E-mail:
  r.roemer@warwick.ac.uk}

\begin{abstract}
  The local density of states (LDOS) of the adsorbate induced
  two-dimensional electron system (2DES) on n-InAs(110) is studied by
  low-temperature scanning tunneling spectroscopy.  In contrast to a
  similar 3DES, the 2DES LDOS exhibits 20 times stronger corrugations
  and rather irregular structures.  Both results are interpreted as a
  consequence of weak localization.  Fourier transforms of the LDOS
  reveal that the $k$-values of the unperturbed 2DES still dominate
  the 2DES, but additional lower $k$-values contribute significantly.
  To clarify the origin of the LDOS patterns, we measure the potential
  landscape of the same 2DES area allowing to calculate the expected
  LDOS from the single particle Schr\"odinger equation and to directly
  compare it with the measured one.
\end{abstract}

%
%
\end{frontmatter}

\section{Introduction and Experimental Details}

Two-dimensional electron systems (2DES) are intensively studied as a
paradigmatic case for many-particle systems in disordered potentials
\cite{2DES}.  They exhibit unique properties with respect to their
three-dimensional counterparts such as weak localization or the
quantum Hall effect \cite{And}.  Many experiments probed the
macroscopic properties of a 2DES, but little is known about the
underlying local density of states (LDOS). On the other hand, detailed
predictions for the LDOS exist from theory \cite{And,Prange} making it
important to establish quantitative LDOS studies \cite{Yama}.

For this purpose, we use the adsorbate induced 2DES \cite{MorKMG02} on
InAs(110) \cite{FeInAs}. In contrast to usual heterostructures
\cite{Ashoori}, this 2DES provides a spatial resolution of $5$ nm well
below characteristic length scales of the 2DES LDOS.  Subband energies
are determined by angle-resolved photoelectron spec\-tro\-scopy
(ARUPS), the disorder potential is measured using the lowest state of
the tip induced quantum dot (QD) \cite{QD}, and the 2DES LDOS is
recorded by scanning tunneling spec\-tro\-scopy (STS).  Thus, for the
first time, all ingredients of the Schr\"odinger equation (SE) are
known and the output of the SE (LDOS) is measured.  We found that the
tendency of a 2DES to weakly localize results in strong and irregular
LDOS corrugations in remarkable contrast to 3DES's, where only weak
and regular corrugations are found \cite{PRL1}.



The UHV-low temperature STM working at $T=6$ K with spectral
resolution in STS down to 0.5 mV is described in \cite{PRL1}.
Degenerate n-InAs ($N_{\rm D} = 1.1 \times 10^{16} / $cm$^{3}$) is
cleaved in-situ at a base pressure of $10^{-8}$ Pa, which leads to a
nearly defect free InAs(110) surface with a Fermi level $E_F=5$ meV
above the conduction band minimum.  To induce the 2DES, Fe is
deposited from an e-beam evaporator \cite{FeInAs}. The Fe coverage is
determined by counting the Fe-atoms and given with respect to the unit
cell of InAs(110).  

Topographic STM-images are recorded in constant current mode with
voltage $V$ applied to the sample. The $dI/dV$-curves are measured by
lock-in technique ($f =$ 1.5 kHz, $V_{mod}=$1.8 mV) with fixed
tip-surface distance stabilized at current $I_{stab}$ and voltage
$V_{stab}$.  The influence of the spatially changing tip-surface
distance is checked to be of minor importance \cite{PRL1}.  

\section{Results and Discussion}
\label{Sec:Res}

\begin{figure}[btp]
\begin{center}\leavevmode
  \includegraphics[width=0.8\linewidth]{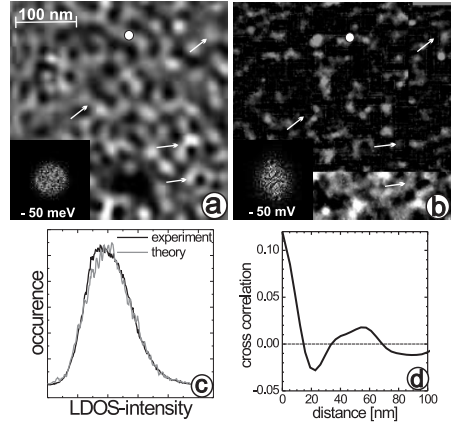}
\caption{\label{Fig4}
  (a) LDOS calculated from a potential landscape at $E=-50$ meV.
  (b) Normalized $dI/dV$-image of the same area; $V=-50$ mV,
  $V_{stab}=100$ mV, $I_{stab}=300$ pA.  Insets are FT's. Dots mark
  identical sample positions as deduced from constant current images.
  (c) Intensity distribution of the LDOS in (a) and (b); for the sake
  of comparison the experimental curve is stretched by 5 \%. 
  (d) Cross correlation function between experimental and calculated
  image. }
\label{figurename}\end{center}\end{figure}

In Fig.~\ref{Fig4}a, we show one of the LDOS images recorded at 2.7 \%
coverage (in the absence of a QD). The spatial resolution ($5$ nm) is
well below the Fermi wave length ($23$ nm).  The total intensity
corresponds to about $30$ electronic states, but, since the scattering
length and thus the localization length is larger than the image size,
more states contribute with part of its intensity.  In general, the
LDOS images exhibit corrugations decreasing in length scale with
increasing $V$ and do not show the circular structures found in the
InAs 3DES \cite{PRL1}. The corrugation strength defined as the ratio
between spatially fluctuating and total $dI/dV$-intensity is $60\pm 5$
\%, i.e., $20$ times larger than the corrugation strength in the 3DES
($3 \pm 0.5$ \%) \cite{PRL1}.

Both results reflect the tendency of the 2DES to weakly localize
\cite{And}. Many different scattering paths containing each many
scattering events contribute to the LDOS leading to more intricate
patterns and the tendency for localization leads to the increased
corrugation.  

A Fourier transform (FT) of the LDOS (inset) reveals the distribution
of contributing $k$-values.  At low voltage a disk is visible in the
FT, which at higher voltage is confined by a ring. At even higher
voltages ($V>-40$ mV), we find that a second smaller disk appears
indicating the occupation of the second subband. We can use these data
to reconstruct the dispersion curve for the lower subband with good
accuracy \cite{MorKMG02}.

We next solve the SE for noninteracting particles numerically using
periodic boundary conditions, the measured disorder potential and {\em
  no} adjustable fit parameter \cite{Metz}.  To construct the LDOS,
the resulting squared wave functions are weighted with the known
energy resolution of the experiment.  The resulting LDOS for a
particular energy is shown in Fig.~\ref{Fig4}a in comparison with the
measured LDOS in Fig.~\ref{Fig4}b.  The correspondence is reasonable,
i.e.  several features as the central ring structure or other smaller
structures marked by arrows appear in both images.  The FT's (insets)
and the intensity distributions of the LDOS (Fig.~\ref{Fig4}c) show
very good agreement demonstrating that the additional $k$-values in
the FT's and the strength of the corrugation are indeed caused by the
interaction with the potential disorder.  Fig. \ref{Fig4}d shows the
cross correlation function between experimental and theoretical
images. Oscillations on the length scale of the unperturbed electron
wave length are found, which demonstrates quantitatively the reasonable
agreement between calculated and measured patterns.

%
%

%
%
\begin{ack}
Financial support from Wi 1277/15-1, the DFG (SPP ``Quanten-Hall-Systeme'', SFB393) and the BMBF
project 05 KS1FKB is gratefully acknowledged.
\end{ack}

%
%


\end{document}